QUANTUM OPTICS

# Push-button photon entanglement

An on-demand source of entangled photons – with one and only one pair each time – has now been generated from a semiconductor chip, which can be a useful resource for optical quantum information.

Chao-Yang Lu and Jian-Wei Pan

--

Quantum entanglement, derided by Einstein as "spooky action at a distance", has entered laboratories as an essential resource for quantum information processing. So far, entanglement between two photons has been produced through parametric down-conversion[1] in a nonlinear optical medium, or radiative cascades in single quantum emitters such as atoms[2] and quantum dots[3,4].

For both fundamental and applied reasons, there is a demanding "quantum wish list" for a perfect entangled photon source: 1. Deterministic generation. Upon a pulsed excitation, the source should emit one and only one pair of entangled photons with a vanishing multi-photon probability; 2. Fidelity. The created two-photon state should closely resembles the target entangled state; 3. Indistinguishability. Individual photons emitted at different trials should be quantum mechanically identical to each other; 4. Collection efficiency. The radiated photons should be extracted with a high efficiency.

However, simultaneously meeting the full list has been challenging. For instance, in parametric conversion – the most widely used workhorse to date – photon pairs are generated only probabilistically, and inevitably accompanied with undesirable multi-pair emissions, posing a serious obstacle to scaling up[5]. In an attempt to overcome this obstacle, increasing attention has turned to single quantum emitters. Writing in Nature Photonics[6], Markus Müller and coworkers describe the first deterministic generation of high-quality two-photon entanglement from a single self-assembled InAs/GaAs quantum dot. Using a clever optical pulsed excitation method[7], they managed to produce, in an on-demand fashion, entangled photons pairs that simultaneously have high single-photon purity, entanglement fidelity and photon indistinguishability, taking one step closer to the wish list.

The semiconductor quantum dots offer an attractive solid state platform for the generation of single and entangled photons. In particular, being compatible with the modern semiconductor device technology, the quantum dots can be artificially engineered through crystal growth and nanofabrication techniques, and monolithically combined with micro- and nano-cavities to tailor light-matter interaction. For over a decade, the most routinely used excitation method was based on non-resonant pumping[3,4] – via above band-gap or *p*-shell. This method is convenient as the emitted single photons can be spectrally separated from the excitation laser background. However, it could cause undesired effects including uncontrolled carrier recombination, emission time jitter, and reduced photon indistinguishability.

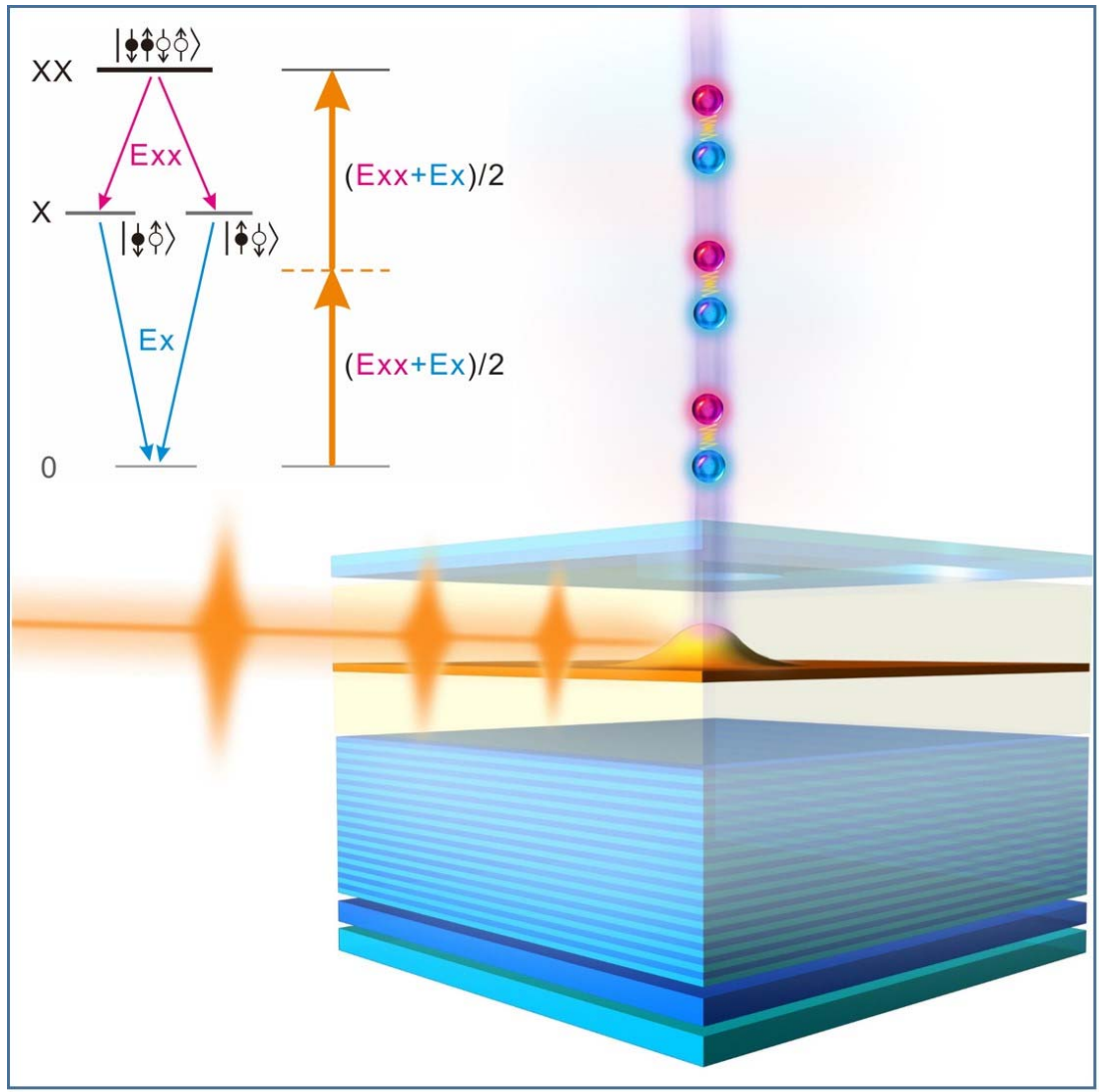

Fig.1: A semiconductor recipe for high-quality photonic entanglement. The top right panel shows the level structure of biexciton-exciton cascade decay. The decay starts from a biexciton (XX) state formed by two electrons and two holes. One of the electrons recombines with one of the holes and generates a first photon, leaving an exciton (X) in the dot which subsequently also recombines to generate a second photon. The bottom left panel illustrates the experimental layout where a pulsed laser excited from the side and the emitted photon pairs are collected from the top. The single quantum dot is embedded in a planar microcavity consisting of 15 lower and 1 upper distributed-Bragg-reflector mirrors.

It is anticipated that *s*-shell resonant excitation, a more controlled method widely used in standard atomic physics experiments, would overcome these shortcomings and yield single photons of better quality. Recent years have witnessed considerable progress along this line. Near background free resonance fluorescence have been obtained from coherently driven quantum dots[8,9], and allowed the observation of textbook examples of Mollow triplet and Rabi oscillation in solid state. In particular, under pulsed *s*-shell excitation, deterministically generated single photons have shown an indistinguishability close to unity[10].

Going beyond single photons, Müller *et al.* have now taken another step forward: generation of entangled photons using coherent excitation[6]. The scheme is based on two-photon emission from biexciton-exciton cascade radiative decay in a single quantum dot (see Fig.1), a semiconductor analog to the seminal work by Aspect *et al.* using Calcium atoms. The polarization of emitted photons is determined by the spin of the intermediate exciton states. To ensure entanglement in the polarization degree of freedom, the fine structure splitting between the two exciton states should be smaller than the radiative linewidth, leaving no room for leaking which-path information.

So the key to the experiment is to deterministically pump the single quantum dot to the biexciton states. It should be noted that the two successively emitted biexciton photon (red arrow in Fig.1) and exciton photon (blue arrow in Fig.1) have distinct energies owing to different charge combinations (thus different Coulomb energies) of their initial and final states. Thus, using single-color laser to climb up the cascade level might be difficult.

Instead, a more elegant excitation method, coherent two-photon excitation[7], has been exploited by Müller and colleagues. The energy of a pulsed laser (yellow arrow in Fig.1) is set at the average energy of the biexciton and exciton photons, in resonance with the virtual biexciton two-photon excitation state. The experimental layout is schematically illustrated in Fig.1. An orthogonal geometry for excitation (side) and collection (top) is used to suppress the laser scattering. An extra bonus from the two-photon excitation scheme is that since the laser energy differs from both the biexciton and exciton photons, the laser background can be spectrally separated, without resorting to polarization suppression technique.

Thanks to the coherent two-photon excitation, a number of criteria in the wish list has now been fulfilled. A clear observation of biexciton Rabi oscillation, together with a numerical analysis, suggests a biexciton preparation efficiency of 98(7)% and photon-pair production efficiency of 86(8)%. Both photons from the pairs are tested by second-order correlation measurements revealing a near-perfect single-photon purity ($g^2(0)<0.4\%$). The entanglement between the two single photons are verified through two-photon correlation in linear, diagonal and circular basis of photon polarization, which yields a state fidelity of 0.81(2), the overlap of the experimentally produced state with the ideal one.

These results have indicated a pretty good source of entangled photons in itself. Furthermore, the authors have shown that these photons are indistinguishable such that they can be further fused together to form entangled states with larger number of photons[5]. This is demonstrated through a two-photon quantum interference experiment, where two identical single photons impinged on a beam splitter coalesce into the same spatial modes. The measured interference visibilities for the biexciton photons and exciton photons are 0.86(3) and 0.71(4), respectively.

Overall, a considerable improvement on the entangled-photon quality has been obtained compared to the results from previous non-resonant excitation methods. Yet, one element still missing from the wish list is the high photon collection efficiency. In their experiment, Müller *et al.* estimated the collection efficiency to be about 0.4%. This figure of merit needs to be substantially increased for scalable photonic quantum computing. High-efficiency extraction of two photons of different color could be achieved through broadband coupling using integrated solid immersion lens[11,12], dielectric planar antennas[13], photonic crystal waveguide[14], nanowires[15], micropillar[16] or "photonic molecular" cavities[17].

With further improvements in the collection efficiency and entanglement fidelity, as well as high-efficiency single-photon detectors and quantum memories, the quantum dot sources of entangled photons could provide a new solid-state platform for optical quantum information[5] where interesting experiments, such as quantum teleportation, entanglement swapping and purification, cluster-state quantum computing and Boson sampling, can be perform with a high efficiency and without worrying about multi-pair emissions.

The authors are at the Hefei National Laboratory for Physical Sciences at the Microscale and Department of Modern Physics, University of Science and Technology of China, Hefei, Anhui, 230026, China.